\documentclass[twocolumn,prb,showpacs,preprintnumbers,amsmath,amssymb]{revtex4}

\usepackage{graphicx}
\usepackage{dcolumn}
\usepackage{bm}
\usepackage{epsfig}

\begin{document}


\title{Mechanical and dielectric relaxation spectra in seven highly viscous glass formers}

\author{U. Buchenau}
\email{buchenau-juelich@t-online.de}
\affiliation{%
Institut f\"ur Festk\"orperforschung, Forschungszentrum J\"ulich\\
Postfach 1913, D--52425 J\"ulich, Federal Republic of Germany}%
\date{April 2, 2007}

\begin{abstract}
Published dielectric and shear data of six molecular glass formers and one polymer are evaluated in terms of a spectrum of thermally activated processes, with the same barrier density for the retardation spectrum of shear and dielectrics. The viscosity, an independent parameter of the fit, seems to be related to the high-barrier cutoff time of the dielectric signal, in accordance with the idea of a renewal of the relaxing entities after this critical time. In the five cases where one can fit accurately, the temperature dependence of the high-barrier cutoff follows the shoving model. The Johari-Goldstein peaks, seen in four of our seven cases, are describable in terms of gaussians in the barrier density, superimposed on the high-frequency tail of the $\alpha$-process. Dielectric and shear measurements of the same substance find the same peak positions and widths of these gaussians, but in general a different weight.
\end{abstract}


\maketitle

\section{Introduction}

The publications of Kia Ngai deal with more substances and more measurement techniques than the work of any other scientist in the field of undercooled liquids. Within the past three decades, whenever a new development appeared, he was the quickest to appreciate it, analyze it and bring it to the general attention, thus speeding up the progress substantially.

Many scientists in the field share his conviction that the flow process in highly viscous liquids can only be understood by combining all possible techniques for its study \cite{birge,ngai,chang,donth,bow,bzow}. The present paper evaluates  recently published \cite{niss,jakobsen} broadband shear and dielectric relaxation data on seven glass formers. The mechanical shear data were obtained with a new technique \cite{christensen} which allows to cover a large dynamical range. Samples for dielectric and shear measurements were taken from the same charge, and the temperature sensors of both measurements were calibrated to each other.

The data show a striking similarity of $G''(\omega)$ and $\epsilon''(\omega)$ on the right hand side of the $\alpha$-peak, a similarity which is sometimes perturbed by the secondary Johari-Goldstein peak \cite{johari1,johari2}. The similarity suggests a common origin of the $\alpha$-peak in dielectrics and shear. The fact that the shear peak appears at a higher frequency than the dielectric peak is explainable in terms of the viscosity, which in a compliance treatment \cite{ferry} is a free parameter. 

In order to identify the elementary processes with thermally activated jumps over an energy barrier $V$, one can use a recent translation \cite{fv} of the textbook \cite{ferry} retardation spectrum $L(\ln\tau)$ into a barrier density function $l(V)$. We will argue that the compliance barrier density of the shear equals the electric dipole moment barrier density $l_\epsilon(V)$ of the dielectric data.

The next section (section II) explains and motivates this approach in more detail. The results of the data treatment are described in section III. They are discussed and compared to other approaches in section IV. Section V summarizes and concludes the paper.

\section{The barrier density functions for shear and dielectrics}

The choice of a retardation spectrum for the shear is motivated by a surprising coincidence, which is more or less visible in the data of all seven substances. We show the two examples where it is most clearly seen in Fig. 1 and Fig. 2.

\begin{figure}[b]
\hspace{-0cm} \vspace{0cm} \epsfig{file=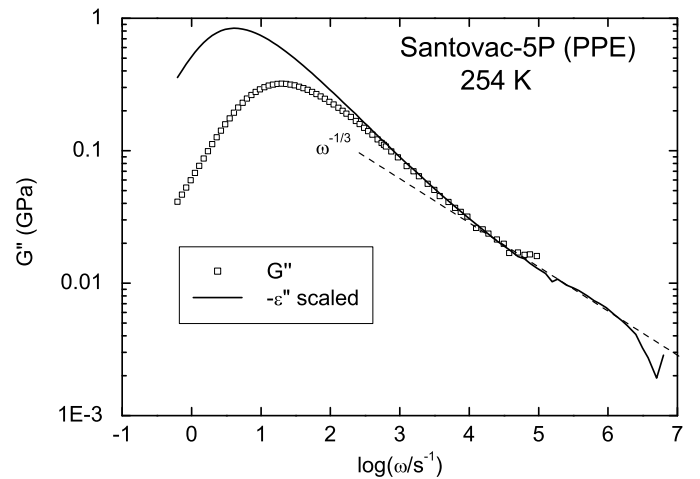,width=9
cm,angle=0} \vspace{0cm} \caption{Comparison of $G''(\omega)$ to a properly scaled $-\epsilon''(\omega)$ in a double-log scale for PPE at 254 K.}
\end{figure}

Fig. 1 compares $G''(\omega)$ and $-\epsilon''(\omega)$ at the same temperature for the type-A glass former PPE. PPE, 1,3-Bis(3-phenoxyphenoxy)benzene, is a diffusion pump oil with the commercial name Santovac-5P, a molecule consisting of five phenyl rings connected by four oxygens to form a short chain. In terms of the classification proposed by the Bayreuth group \cite{kudlik}, it is a type-A glass former, a glass former which shows no or at least no pronounced secondary Johari-Goldstein peak. The (negative) $\epsilon''(\omega)$-data have been scaled to coincide with the $G''(\omega)$-data on the right hand side of the peak.  One finds good agreement between $G''(\omega)$ and $-\epsilon''(\omega)$ as soon as the frequency is two decades higher than the one of the peak in $G''(\omega)$.

In addition, Fig. 1 shows a change of slope of the $\alpha$-tail, from an $\omega^{-1/2}$- to an $\omega^{-1/3}$-behavior. The tendency is seen most clearly in the dielectric data, but it is also in the shear data; their fit improves markedly if one allows for a $\omega^{-1/3}$-component. This might be the influence of a hidden Johari-Goldstein peak, an explanation which has been favored for glycerol on the basis of pressure, aging and chemical series measurements (for a review, see Ngai and Paluch \cite{paluch}). But there is also impressive experimental evidence for a limiting $\omega^{-1/3}$-behavior in shear compliance data of type-A molecular glass formers \cite{plazek}, which show the so-called Andrade \cite{andrade} creep, $J(t)\sim t^{1/3}$, in the short-time limit. Therefore we will fit our data in terms of a sum of an $\omega^{-\beta}$-term (with $\beta$ as fit parameter) and an $\omega^{-1/3}$-term, dominating at high frequency.

Toward lower frequency, the common slope terminates for the shear data already at a higher frequency (a shorter time) than for the dielectric data. The natural explanation for this is that the parallel between dielectrics and shear is in fact between shear compliance and dielectric susceptibility. In a comparison of these two quantities, the shear compliance starts to deviate from the dielectric susceptibility as soon as the viscous flow sets in. This suggests a treatment of the shear data in terms of a retardation spectrum, with the viscosity as an independent parameter \cite{ferry}.  

\begin{figure}[b]
\hspace{-0cm} \vspace{0cm} \epsfig{file=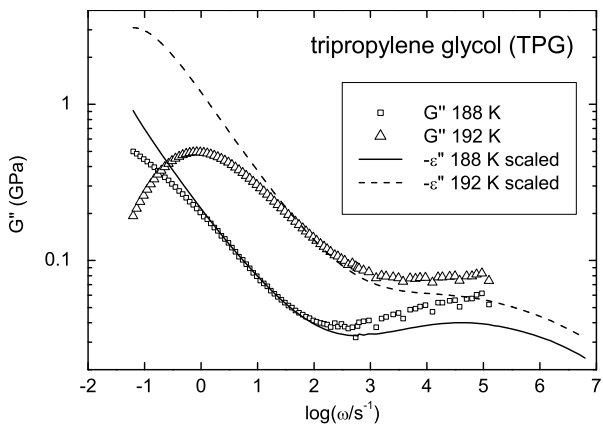,width=8
cm,angle=0} \vspace{0cm} \caption{Comparison of $G''(\omega)$ to a properly scaled $-\epsilon''(\omega)$ in a double-log scale for TPG at 188 and 192 K.}
\end{figure}

The good agreement at the right-hand side of the $\alpha$-peak is found to be a general feature of all seven measured substances, as long as there is no disturbing influence from the secondary Johari-Goldstein peak \cite{johari1,johari2}. This is seen in Fig. 2, which shows that the good agreement between $G''$ and $-\epsilon''$ disappears as the peaks merge. The substance, tripropylene glycol (TPG), $C_6H_{20}O_4$, is still a molecule and not yet a polymer (the whole series from the small molecule propylene glycol to long-chain polypropylene glycol is well-investigated by dielectrics \cite{leon,lars1,lars2}). TPG itself has been studied under aging \cite{dyretpg} and under pressure \cite{pawlus}. It consists of three connected propylene groups, with a large dielectric moment and a pronounced Johari-Goldstein peak (which is absent or at least much less pronounced in propylene glycol \cite{leon}).

Fig. 2 shows another tendency which will be evaluated quantitatively in this paper, namely a much stronger increase of the imaginary quantities in the $\alpha$-peak region with temperature than in the Johari-Goldstein peak region.

If one wants to decompose a measured relaxation into a spectrum of exponential decays in time, one can choose between two equivalent possibilities \cite{ferry}, the relaxation spectrum in which the elementary exponential relaxators add to decrease the modulus or the retardation spectrum in which they add to increase a susceptibility. In principle, the choice is not crucial, because the two spectra can be calculated from each other. Here, we choose the retardation spectrum, with the viscosity $\eta$ as an independent variable.

For this choice, one has the textbook expressions \cite{ferry} for the real and imaginary parts of the complex frequency-dependent shear compliance
\begin{equation} \label{jpferry}
J'(\omega)=J_g+\int_{-\infty}^\infty L(\ln\tau)\frac{1}{1+\omega^2\tau^2} d(\ln\tau)
\end{equation}
and
\begin{equation}\label{jppferry}
J''(\omega)=\int_{-\infty}^\infty L(\ln\tau)\frac{\omega\tau}{1+\omega^2\tau^2} d(\ln\tau)+\frac{1}{\omega\eta},
\end{equation}
where $\tau$ is the relaxation time and $L(\ln\tau)$ is the weight of this relaxation time in the retardation spectrum. $J_g$, the glass compliance, is the inverse of the infinite frequency modulus $G_\infty$.

In an energy landscape picture \cite{goldstein}, one reckons with thermally activated jumps over the energy barrier between two neighboring minima. In fact, one very often finds a broad secondary relaxation peak (the Johari-Goldstein peak \cite{johari1,johari2}) below the $\alpha$-peak of the flow process. This peak follows the Arrhenius relation in the glass phase, indicating that it stems from local thermally activated jumps. For a jump over an energy barrier of height $V$, the Arrhenius relation for the relaxation time $\tau_V$ reads
\begin{equation}
\tau_V=\tau_0{\rm e}^{V/k_BT},
\end{equation}
where $\tau_0=10^{-13}$ s and $T$ is the temperature.

For a spectrum of thermally activated jumps, one defines \cite{fv} the barrier density function $l_s(V)$
\begin{equation}
l_s(V)=\frac{G_\infty}{k_BT} L(V/k_BT+\ln\tau_0).
\end{equation}
The index $s$ stands for the shear. With this definition, the complex shear compliance equations (\ref{jpferry}) and (\ref{jppferry}) transform into
\begin{equation}\label{jp}
J'(\omega)=J_g+J_g\int_0^\infty l_s(V)\frac{1}{1+\omega^2\tau_V^2} dV
\end{equation}
and
\begin{equation}\label{jpp}
J''(\omega)=J_g\int_0^\infty l_s(V)\frac{\omega\tau_V}{1+\omega^2\tau_V^2} dV+\frac{1}{\omega\eta}.
\end{equation}

The dielectric susceptibility can also be described \cite{pollak,gilroy} in terms of a dielectric barrier density function $l_\epsilon(V)$  
\begin{equation}\label{ep}
\frac{\epsilon'(\omega)-\epsilon_\infty}{\epsilon(0)-\epsilon_\infty}=\int_0^\infty l_\epsilon(V)\frac{1}{1+\omega^2\tau_V^2} dV
\end{equation}
and
\begin{equation}\label{epp}
\frac{\epsilon''(\omega)}{\epsilon(0)-\epsilon_\infty}=\int_0^\infty l_\epsilon(V)\frac{\omega\tau_V}{1+\omega^2\tau_V^2} dV.
\end{equation}
Here $\epsilon(0)$ is the static dielectric susceptibility, $\epsilon_\infty$ is the real part of $\epsilon(\omega)$ in the GHz range (larger than $n^2$, the square of the refractive index, because of vibrational contributions \cite{bzow}).

The above definitions of eqs. (\ref{jp}-\ref{epp}) imply a normalization of both $l_s(V)$ and $l_\epsilon(V)$ with
\begin{equation}\label{snorm}
	\int_0^\infty l_s(V)dV=\frac{J_e^0-J_g}{J_g},
\end{equation}
where $J_e^0$ is the recoverable compliance of the steady-state flow \cite{ferry}, and
\begin{equation}\label{enorm}
	\int_0^\infty l_\epsilon(V)dV=1.
\end{equation}

The dielectric $\alpha$-peak occurs always at a lower frequency than the shear one and seems to coincide with the heat capacity and the structural relaxation peaks \cite{birge,ngai,chang,donth,bow,bzow}. Below, we will adopt the view that the left side of the dielectric peak marks the disappearance and renewal of the relaxing entities.

In order to describe this decay, one needs to multiply the barrier density of the energy landscape with an appropriate cutoff function at a cutoff barrier $V_c$. Here, we will assume that the relaxing entities decay exponentially in time with the critical relaxation time $\tau_c$. With the Arrhenius relation $\tau_c=\tau_0\exp(V_c/k_BT)$, this translates into a double-exponential cutoff
\begin{equation}
c(V)=\exp(-\exp((V-V_c)/k_BT)).
\end{equation}

\begin{figure}[b]
\hspace{-0cm} \vspace{0cm} \epsfig{file=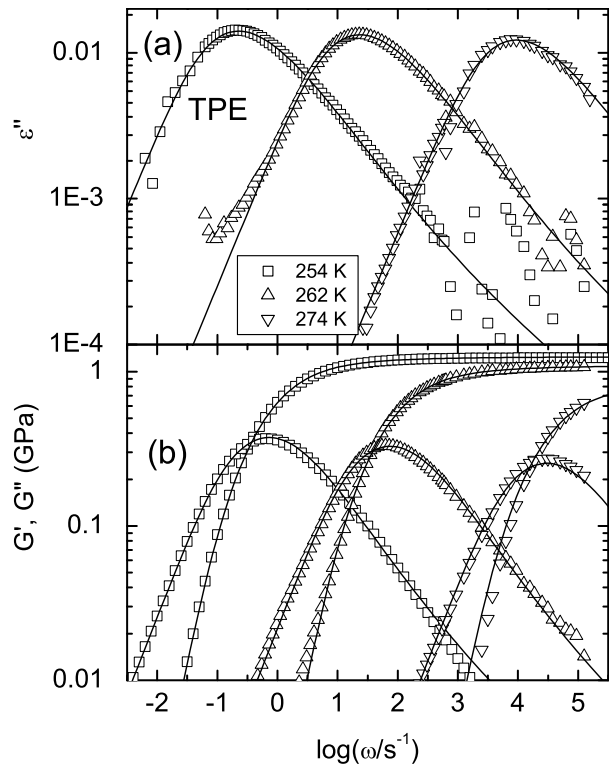,width=8
cm,angle=0} \vspace{0cm} \caption{(a) Data and fit of $\epsilon''(\omega)$ in a double-log scale for TPE between 254 and 274 K (b) the same for $G(\omega)$.}
\end{figure}

Equations (\ref{jpp}) and (\ref{epp}) show that a Johari-Goldstein peak in $G''(\omega)$ or $\epsilon''(\omega)$ at the peak frequency $\omega_1$ corresponds to a peak in $l(V)$ at a peak barrier $V_1=k_BT\ln(1/\omega_1\tau_0)$. We will see that the Johari-Goldstein peaks are reasonably well described by gaussians in $l(V)$.

\begin{figure}[b]
\hspace{-0cm} \vspace{0cm} \epsfig{file=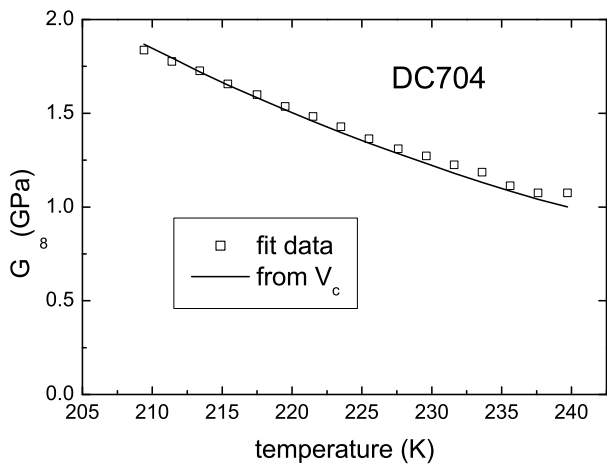,width=8
cm,angle=0} \vspace{0cm} \caption{Fit values of $G_\infty$ in DC704 as a function of temperature. The continuous line is the temperature dependence $V_c/\Delta v$ (with $\Delta v=0.057\ nm^3$) expected from the shoving model \cite{shoving}.}
\end{figure}

To describe both the $\alpha$-peak and the Johari-Goldstein-peak in terms of a barrier density, $l_s(V)$ and $l_\epsilon(V)$ will be fitted by the form
\begin{equation}\label{lvform}
l(V)=(a_\beta {\rm e}^{\beta V/k_BT}+a_{1/3}{\rm e}^{V/3k_BT}+a_1{\rm e}^{-\gamma_1(V-V_1)^2})c(V).
\end{equation}
The first two terms describe the high-frequency tail of the $\alpha$-process, the third term the Johari-Goldstein peak (if there is one; in three of our seven examples, it is not needed).  The dimensionless parameter $\beta$ determines the slope $\omega^{-\beta}$ at the beginning of the $\alpha$-tail in the double-log plot of Fig. 1.

Instead of using the three prefactors $a_\beta$, $a_{1/3}$ and $a_1$ as fit parameters, it is better to use the corresponding weights $w_\beta$, $w_{1/3}$ and $w_1$ in the integral over the barriers, equs. (\ref{snorm}) and (\ref{enorm}). A type-A glass former without Johari-Goldstein peak with $w_1=0$ is characterized by the two dimensionless parameters $\beta$ and $b_2=w_{1/3}/(w_\beta+w_{1/3})$, at least as far as the form of its spectrum is concerned. $\beta$ and $b_2$ have the advantage to be reasonably temperature-independent.

With this prescription, one can fit the $\epsilon''(\omega)$ of a type-A glass former with two temperature-independent parameters, $\beta$ and $b_2$, and two temperature-dependent parameters, $\Delta\epsilon=\epsilon(0)-\epsilon_\infty$ and $V_c$. Their temperature dependence is a decrease with increasing temperature, which is well fitted by an appropriate power law
\begin{equation}\label{tscal}
	\Delta\epsilon(T)=\Delta\epsilon(T_g)\left(\frac{T_g}{T}\right)^{\gamma_\epsilon},
\end{equation}
where $T_g$ is the glass temperature. Similarly, one describes the decrease of $V_c$ with the exponent $\gamma_V$ and the one of $G_\infty$ with $\gamma_G$.

The strategy of our evaluation is to fit $l(V)$ to the dielectric data, and then use the same spectral form to describe the shear. The fit of the shear data requires three additional temperature-dependent parameters, $J_g$, $J_e^0$ and $\eta$. Again, it is worthwhile to look for combinations which might turn out to be temperature-independent. One of them is the ratio
\begin{equation}\label{f0}
	f_0=\frac{J_e^0-J_g}{J_g},
\end{equation}
which appears in the normalization of the shear spectrum, eq. (\ref{snorm}). A second interesting possibility is not to fit the viscosity $\eta$, but the ratio
\begin{equation}\label{fjc}
	f_{jc}=\frac{\tau_c}{f_0J_g\eta},
\end{equation}
where $\tau_c$ is the Arrhenius relaxation time of the terminal barrier $V_c$. As we will show in the discussion, one can argue that the ratio $f_{jc}$ should be 2 for a renewal of the relaxing entities within the critical time $\tau_c$.

In the case of a type-B glass former, Fig. 2 shows that one needs another dimensionless parameter, because the weight of the Johari-Goldstein peak is different in the two quantities. In Fig. 2, the Johari-Goldstein peak is more prominent in the shear signal, but this varies from substance to substance.

\section{Data evaluation}

\subsection{The three type-A glass formers}

Three of our seven substances, TPE, DC704 and PPE, happen to have no or at least only a rather weak Johari-Goldstein peak. Let us begin with TPE. TPE stands for triphenylethylene, $C_{20}H_{16}$, a rather flexible molecule with three phenyl rings attached to a central $C=C$ double bond. Fig. 3 (a) shows data and fit for $\epsilon''(\omega)$ in a double-log plot, Fig. 3 (b) the ones for $G(\omega)$. The dielectric data in Fig. 3 (a) are well fitted with only the first two terms of eq. (\ref{lvform}), without any Johari-Goldstein peak. $\beta$ and $b_2$ turn out to be temperature-independent within experimental accuracy.

One gets a good fit for the shear data in Fig. 3 (b), taking over $\beta$, $b_2$ and the cutoff barrier $V_c$ from the fit of the dielectric data at the given temperature and fitting $G_\infty$, $f_0$ and $f_{jc}$. $G_\infty$ is temperature-dependent, but $f_0$ and $f_{jc}$ are again temperature-independent within the experimental accuracy, thus justifying our choice of variables. The parameters and their temperature dependence are listed in Table I.

The temperature exponents $\gamma_V$ and $\gamma_G$ of the critical barrier $V_c$ and the infinite frequency shear modulus $G_\infty$ are the same within their error bars (about 5 \% for $\gamma_V$ and about 10 \% for $\gamma_G$). This shows the validity of the shoving model \cite{shoving}, according to which the energy barrier of the $\alpha$-process should be proportional to the infinite frequency shear modulus $G_\infty$. The shoving model postulates that the $\alpha$-process happens when the local energy concentration exceeds the product $G_\infty\Delta_v$, where $\Delta_v$ is a volume expansion.

The same results, maybe even a bit clearer because of the stronger dielectric signals, are obtained for the two other type-A glass formers PPE and DC704. Again, the fit parameters are listed in Table I. In particular, the $\omega^{-1/3}$-contribution is much better seen, as illustrated in Fig. 1 for PPE. In DC704, again a diffusion pump oil (1,3,3,5-tetramethyl-1,1,5,5-tetraphenyl-trisiloxane, a rather large molecule) we have the additional advantage of a large temperature range of the measurement, from 209 to 239 K. As in TPE, we find temperature-independent parameters $\beta$, $b_2$, $f_0$ and $f_{jc}$. Again, we find the shoving model \cite{shoving} confirmed in both glass formers. In DC704, one even sees the curvature of both curves (see Fig. 4), which justifies our temperature exponent Ansatz, eq. (\ref{tscal}).

\begin{figure}[b]
\hspace{-0cm} \vspace{0cm} \epsfig{file=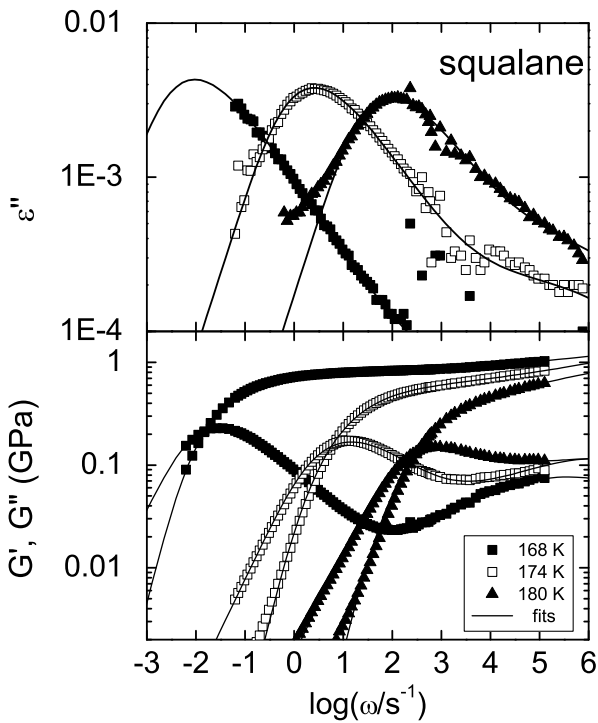,width=8
cm,angle=0} \vspace{0cm} \caption{Data and fits of (a) $\epsilon''(\omega)$ (b) $G(\omega)$ in squalane.}
\end{figure}

\begin{figure}[b]
\hspace{-0cm} \vspace{0cm} \epsfig{file=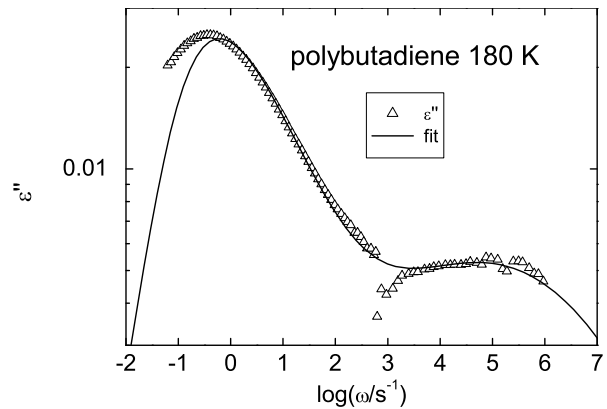,width=8
cm,angle=0} \vspace{0cm} \caption{Data and fit of $\epsilon''(\omega)$ in polybutadiene at 180 K.}
\end{figure}

\begin{figure}[b]
\hspace{-0cm} \vspace{0cm} \epsfig{file=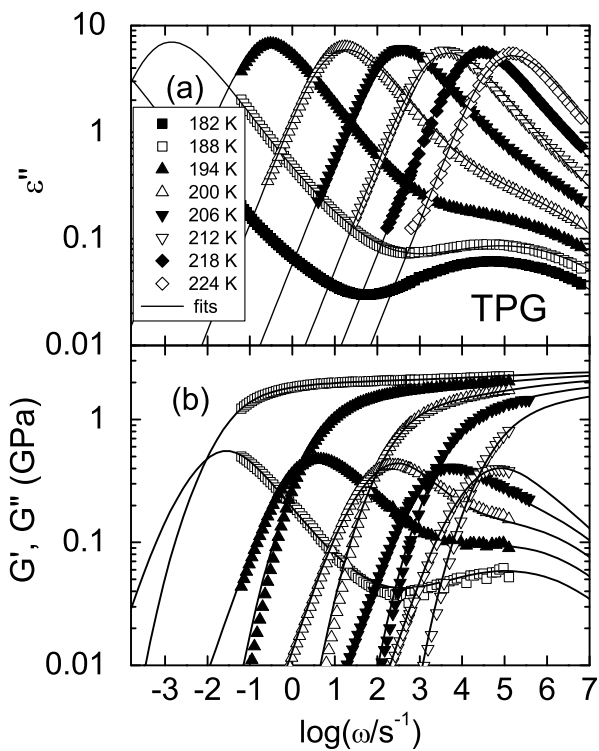,width=8
cm,angle=0} \vspace{0cm} \caption{Data and fits of (a) $\epsilon''(\omega)$ (b) $G(\omega)$ in TPG.}
\end{figure}

Table I comprises the fit parameters for these three type-A glass formers. Note that our formalism allows to describe both shear and dielectric data over the whole temperature range with eleven temperature-independent parameters.

\begin{table}[h]
	\centering
		\begin{tabular}{|c|c|c|c|}
  \hline
  glass former          & TPE      & DC704   & PPE     \\  \hline
$T_g$ (K)               & 249      & 211     & 244     \\  \hline
$\Delta\epsilon$        & 0.0491   & 0.257   & 2.011   \\
$\gamma_\epsilon$       & 1.85     & 2.26    & 1.90    \\ 
$\beta$                 & 0.77     & 0.85    & 1.04    \\ 
$b_2$                   & 0.18     & 0.27    & 0.215   \\
$V_c(T_g)$ (eV)         & 0.767    & 0.639   & 0.755   \\
$\gamma_V$              & 4.3      & 4.6     & 4.6     \\  \hline
$G_\infty(T_g)$ (GPa)   & 1.38     & 1.80    & 1.27    \\
$\gamma_G$              & 4.7      & 4.2     & 4.7     \\ 
$f_0$                   & 1.65     & 2.38    & 2.22    \\ 
$f_{jc}$                & 2.5      & 2.45    & 2.05    \\  \hline
		\end{tabular}
	\caption{Parameters of the three type-A glass formers. Upper part $\epsilon(\omega)$, lower part $G(\omega)$.}
	\label{tab:tab1}
\end{table}

\subsection{The four type-B glass formers}

In the type-B glass formers DHIQ, PB20, Squalane and TPG, one needs to fit a Johari-Goldstein peak on top of the high-frequency tail of the $\alpha$-process. This is illustrated in Fig. 5 for our first type-B example, squalane.

Squalane, $C_{30}H_{62}$, is a short chain molecule with 24 carbon atoms in the backbone and 6 attached $CH_3$-groups, rather polymerlike. It has a strong and well-separated Johari-Goldstein peak (see Fig. 5), much better visible in the shear data than in the dielectric data. The dielectric dipole moment is very weak. Nevertheless, it is possible to fit both sets of data with the same retardation spectrum, attaching a substantially higher weight to the Johari-Goldstein peak in the shear (see Table II).

\begin{figure}[b]
\hspace{-0cm} \vspace{0cm} \epsfig{file=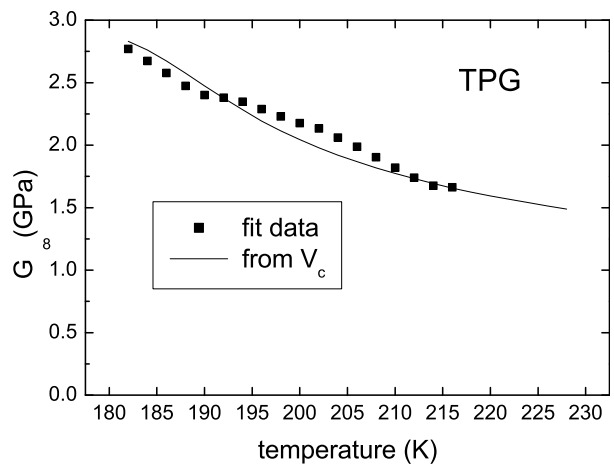,width=8
cm,angle=0} \vspace{0cm} \caption{Fit values of $G_\infty$ in TPG as a function of temperature. The continuous line is the temperature dependence $V_c/\Delta v$ (with $\Delta v=0.037\ nm^3$) expected from the shoving model \cite{shoving}.}
\end{figure}

In this substance, it is not possible to fit the shear data with a temperature-independent parameter $f_0$; one has to postulate a rather strong increase of $f_0$ with increasing temperature
\begin{equation}\label{f0t}
	f_0(T)=f_0(T_g)+f_0'(T-T_g),
\end{equation}
but one can keep the parameter $f_{jc}$ constant (see Table II).

PB20 is a true polymer, relatively short (5000 g/mol), composed of 80 \% 1,4-polybutadiene monomers and 20 \% 1,2-polybutadiene monomers. The results look very similar to those of squalane, and the resulting fit parameters in Table II are in fact close to those of squalane. Even more than squalane, it has the polymer feature of a relatively slow decrease of $\epsilon''(\omega)$ at low frequency, explainable in terms of chain modes with long relaxation times \cite{ferry}. This is illustrated in Fig. 6, which shows the deviation between fit and data at low frequency. As a consequence, the resulting parameters have a larger error bar in squalane and polybutadiene than in the two molecular substances TPG and DHIQ. In particular, the deviations between $\gamma_V$ and $\gamma_G$ do not demonstrate a failure of the shoving model.

TPG is a much more favorable case, with a very large dipole moment and no problems at the cutoff barrier. As Fig. 7 (a) shows, our spectrum of eq. (\ref{lvform}) provides beautiful fits over a large temperature range. One needs to take the temperature dependence of the Johari-Goldstein peak position $V_1$ into account. Our fit found
\begin{equation}\label{tpgshift}
	V_1=0.295\frac{T}{T_g},
\end{equation}
a bit smaller shift than the one found in aging experiments \cite{dyretpg}.

Fig. 7 (b) shows that the shear data are well described in terms of the dielectric retardation spectrum. There is a small temperature dependence of $f_0$, but $f_{jc}$ is again a temperature-independent constant. The shoving model is found to be well fulfilled (see Fig. 8). 

Finally, DHIQ, decahydroisoquinoline, $C_9H_{17}N$, is best described as two cyclohexanol rings sharing one $C-C$-bond, one of the two rings having an $NH$ replacing a $CH_2$-group. In this case, the Johari-Goldstein peak is very prominent in the dielectric data \cite{ranko,paluch2}, comparable to the one in $G(\omega)$. The dipole moment is large; both $V_c$ and $G_\infty$ can be determined with high accuracy. Again, their temperature exponents $\gamma_V$ and $\gamma_G$ agree within the error bars (see Table II), in agreement with the shoving model \cite{shoving}. Since both are exceptionally large (DHIQ is very fragile, m=158 in Angell's scheme \cite{bohmer}), their good agreement provides a strong argument for the validity of the model.

In Table II, the Johari-Goldstein peak is characterized by the weight of the peak
\begin{equation}
	w(T)=a_1\sqrt(\pi/\gamma_1)=a_1FWHM\sqrt(\pi/4\ln 2)
\end{equation}
which shows a Boltzmann factor behavior
\begin{equation}\label{boltz}
w(T)=w(T_g)\exp(-E_a(1/k_BT-1/k_BT_g)),
\end{equation}
with a  formation energy $E_a$ which is on the average 2/3 of the peak position $V_1$.

\begin{table}[h]
	\centering
		\begin{tabular}{|c|c|c|c|c|}
  \hline
  glass former             & Squalane & PB20    & TPG    &  DHIQ   \\  \hline
$T_g$ (K)                  & 167      & 176     & 184    &  175    \\  \hline
$\Delta\epsilon(T_g)$      & 0.0155   & 0.132   & 23.3   &  1.707  \\
$\gamma_\epsilon$          & 2.1      & 0.0     & 1.51   &  0.0    \\ 
$\beta$                    & 0.6      & 0.44    & 0.85   &  0.4    \\ 
$b_2$                      & 0.2      & 0.2     & 0.22   &  0.2    \\
$V_c(T_g)$ (eV)            & 0.517    & 0.54    & 0.63   &  0.635  \\
$\gamma_V$                 & 3.2      & 3.8     & 3.0    &  6.4    \\  \hline
$G_\infty(T_g)$ (GPa)      & 1.33     & 1.63    & 2.69   &  3.1    \\
$\gamma_G$                 & 2.5      & 2.7     & 2.8    &  6.3    \\ 
$f_0(T_g)$                 & 2.4      & 3.25    & 6.7    &  1.56   \\ 
$f_0'$ (1/K)               & 0.4      & 0.41    & -0.04  &  0.25   \\
$f_{jc}$                   & 2.7      & 2.4     & 2.5    &  2.0    \\  \hline
$V_1$ (eV)                 & 0.27     & 0.28    & 0.32*  &  0.32   \\
$FWHM$(eV)                 & 0.135    & 0.170   & 0.154  &  0.16   \\ 
$w_s(T_g)$                 & 0.56     & 0.55    & 0.15   &  0.50   \\ 
$w_\epsilon(T_g)$          & 0.03     & 0.23    & 0.02   &  0.48   \\
$E_a$ (eV)                 & 0.25     & 0.12    & 0.19   &  0.25   \\  \hline
*average value, see eq. (\ref{tpgshift})  &          &         &        &         \\  \hline

		\end{tabular}
	\caption{Parameters of the four type-B glass formers. Upper part $G(\omega)$, middle part $\epsilon(\omega)$, lower part Johari-Goldstein peak parameters for both.}
	\label{tab:tab2}
\end{table}

\section{Discussion}

The preceding section presented a quantitative description of the $\alpha$- and the $\beta$-process in dielectrics and shear for seven different glass formers, a description which is based on the concept of isolated and independent thermally activated jumps in the energy landscape. The description allows for a reasonable fit of the temperature dependence in terms of temperature-independent parameters. The number of parameters is not small; one needs eleven or twelve parameters for a type-A glass former (depending on whether $f_0$ is temperature-independent or not, see Table I and II) and five additional parameters for the description of the $\beta$- or Johari-Goldstein peak (see Table II).

Nevertheless, the exercise is not completely meaningless. One does indeed get meaningful quantitative information, which is impossible to obtain otherwise. The first and rather important one is the probable equality of the retardation spectra of shear and dielectrics (but with a different weight of the Johari-Goldstein peak), an information which one can guess from the raw data (see Figs. 1 and 2), but which requires a full fit for its quantitative check.

The second and equally important quantitative information concerns the dimensionless ratio $f_{jc}$ between the terminal dielectric relaxation time $\tau_c$ and the product of the viscosity with the total retardation compliance, eq. (\ref{fjc}). The seven fitted values lie between 2 and 2.7 (average value 2.37). This indicates a general relation between the dielectric terminal time and the viscosity. Since the dielectric terminal time seems to coincide with the structural lifetime \cite{bow}, it is probably also the lifetime of the double-well potentials which are responsible for the retardation spectrum.

Question: What do we expect for the ratio $f_{jc}$ if this is indeed the case? To answer this question, consider a constant applied shear stress. After the time $\tau_c$, all the double-well potentials of the spectrum would have reached thermal equilibrium, giving their full contribution to the compliance. From this consideration, if we renew them at the time $\tau_c$, we would naively expect them to be able to give their contribution again after this time, yielding $f_{jc}=1$.

But this answer is not correct. To get the correct answer, one must consider the difference between energy and free energy in these double-well potentials. To keep the argument simple, let us restrict ourselves to the special case of a symmetric double-well; it applies as well to the asymmetric case.

If the double-well is initially symmetric and if it couples to the shear stress $\sigma$ with a coupling constant $v$ (the coupling constant has the dimension of a volume), then the asymmetry $\Delta$ under the stress is $\sigma v$. One well has the energy $-\sigma v/2$, the other has the energy $+\sigma v/2$.

In thermal equilibrium, the population of the two wells is given by their Boltzmann factors. It is easy to calculate the energy $U$ of the equilibrated system in the limit of a small stress
\begin{equation}
	U=-\frac{\sigma^2v^2}{4k_BT}.
\end{equation}
This is the energy transported to the heat bath in the equilibration of the relaxing entity after switching on the stress.

The free energy $F$ is
\begin{equation}
	F=-\frac{\sigma^2v^2}{8k_BT},	
\end{equation}
only half of the energy itself. If one thinks about it, the reason is clear: spending the energy, one has spanned an entropic spring by the population difference in the two wells. If one removes the stress slowly, one gets half the energy back. But if the double-well potential decays, one gets nothing back.

The contribution of the relaxing entity to the compliance is given by the second derivative of the free energy with respect to the stress. But if we now deal with the effect of a renewal of the double-well potential on the viscosity, we have to count the energy. This means we spend twice as much energy under a constant stress as the one calculated above in our first oversimplified picture. And this means the viscosity must be a factor of 2 smaller, which implies $f_{jc}=2$. This is reasonably close to the fitted values in Table I and II.

A third quantitative conclusion of the present study is a surprising agreement with the conclusions of Plazek et al \cite{plazek} from their recoverable shear compliance experiments. If one takes the parameters of Table I to calculate the recoverable compliance, one gets curves which closely resemble those reported by them. Obviously, it is experimentally much easier to detect the Andrade creep \cite{andrade} in creep experiments than in dynamical ones.
If one calculates $f_0$ from their data, one finds values between 1.5 and 2.3, similar to those in Table I.

Here, however, a word of caution is in place. Our data, taken as they are, do not imply a limited recoverable shear compliance. In fact, they are well fitted by the BEL model \cite{bel}, which has a divergent recoverable compliance. The values in the two tables stem from the assumption that the two retardation spectra of dielectrics and shear (at least as far as the $\alpha$-peak is concerned) are the same.

The same is true for the fourth conclusion, the validity of the shoving model \cite{shoving}. The fitted $G_\infty$-values were obtained under the same assumption.

Finally, the Johari-Goldstein peak increases its height with increasing temperature. The increase follows a Boltzmann factor, with a formation energy of about two thirds of the barrier height at the center of the peak.

\section{Summary and conclusions}

Dielectric and shear relaxation data in seven highly viscous liquids, most of them molecular liquids, were evaluated in terms of a barrier density of independent thermally activated relaxation centers. Three of the substances are type-A glass formers without or with only a very small Johari-Goldstein peak, four of them show a pronounced Johari-Goldstein peak.

The most important conclusion is the probable equality of the dielectric and shear retardation spectra, guessed from the raw data and confirmed by a quantitative fit. The difference in the peak positions is due to the influence of the viscosity. The Johari-Goldstein peak has different weight in dielectrics and shear. 

The second important conclusion concerns the viscosity. It seems probable that the viscosity results from the constant renewal of the double-well potentials in the sample within the terminal dielectric relaxation time.

Our data support earlier recovery compliance results by Plazek et al \cite{plazek}, according to which one has an Andrade \cite{andrade} creep $J\sim t^{1/3}$ at short times in type-A glass formers (glass formers without Johari-Goldstein peak). They further support the shoving model \cite{shoving}, which postulates a proportionality between the infinite frequency shear modulus and the Arrhenius barrier of the terminal relaxation time.

Acknowledgement: The author is deeply thankful to Kristine Niss and Bo Jakobsen for communicating their beautiful data to him, to Niels Boye Olsen and Tage Christensen for enlightening discussions and to Jeppe Dyre for constant encouragement and a lot of helpful advice.

\end{document}